# Dependence of SWNT Growth Mechanism on Temperature and Catalyst Particle Size: Bulk versus Surface Diffusion

**Feng Ding[*, a], Arne Rosén[a] and Kim Bolton[a, b]**

[a] Department of Physics, Göteborg University, SE-412 96, Göteborg, Sweden

[b]School of Engineering, University College of Borås, SE-50190, Borås, Sweden

The vapor-liquid-solid (VLS) model is often used to explain the growth of carbon nanotubes.[1] In this model the liquid catalyst particle acts i) as a catalyst for the decomposition of the carbon feedstock (e.g., CO, $CH_4$ or alcohols), ii) as a solvent for the carbon atoms that are released from the feedstock and iii) as a template for the nucleation and growth of the carbon nanotubes. However, recent experiments show direct evidence that carbon fibers can grow from solid 6-20 nm catalytic particles,[2] indicating that the particles do not need to be in the liquid phase for growth of one-dimensional carbon structures. It was suggested that carbon atoms diffuse on the surface of these clusters, before they are incorporated in the fiber structure at step defects on the particle surface. Also, single-walled carbon nanotubes (SWNTs), multi-walled carbon nanotubes (MWNTs) and carbon nanofibers (CNFs) can be produced at temperatures much lower the eutectic point of the bulk catalyst [3-7], and the melting point of the (rather large) catalyst particles used in the experiments are lower than the bulk melting points by only 5-10% [8].

---

[*] Corresponding author. fengding@fy.chalmers.se

The metal particle does not need to be in the liquid phase for the catalytic decomposition of carbon feedstock, but it allows for rapid diffusion of carbon atoms into the metal cluster before these atoms are incorporated into the growing SWNT.  In this contribution, we report results of a molecular dynamics (MD) study comparing the nucleation of SWNTs from liquid and solid catalyst particles.

We employ the same potential energy surface (PES) used in previous MD studies of SWNT growth from liquid Fe particles [9-11]. In addition to yielding a growth mechanism that is in agreement with experiment,[9] this PES also yields that correct dependence of the cluster melting point on its size,[8] as well as the correct trends in the FeC phase diagram [8] (which is necessary to allow for possible C precipitation at the eutectic temperature).

Two types of trajectory conditions were considered in this study. The first mimics catalysed carbon vapour deposition (CCVD) experiments with floating catalyst particles (and hence ignores possible substrate effects [4]). Briefly, $Fe_{300}$ clusters were annealed to the desired temperature, after which C atoms were randomly inserted to the surface layer of the cluster (i.e., the atoms are inserted at a distance $d$ from the center of mass of the cluster, where $R<d<1.1R$ and $R=1.0$ nm is the radius of the $Fe_{300}$ cluster). The rate of C deposition on the surface is 1 atom per 50 ps, which is more than four orders of magnitude larger than the upper limit of the CCVD experiments, [12] but is required for the simulation of SWNT nucleation within a reasonable computational time.[13]

The second set of trajectory conditions sheds light on the thermal properties and surface diffusion of the iron-carbide (FeC) clusters.  Here, an annealed $Fe_{300}C_{60}$ cluster was heated from 500 K to 1300 K in steps of 50 K, and the Lindemann index of each atom was determined at each

temperature. [8]

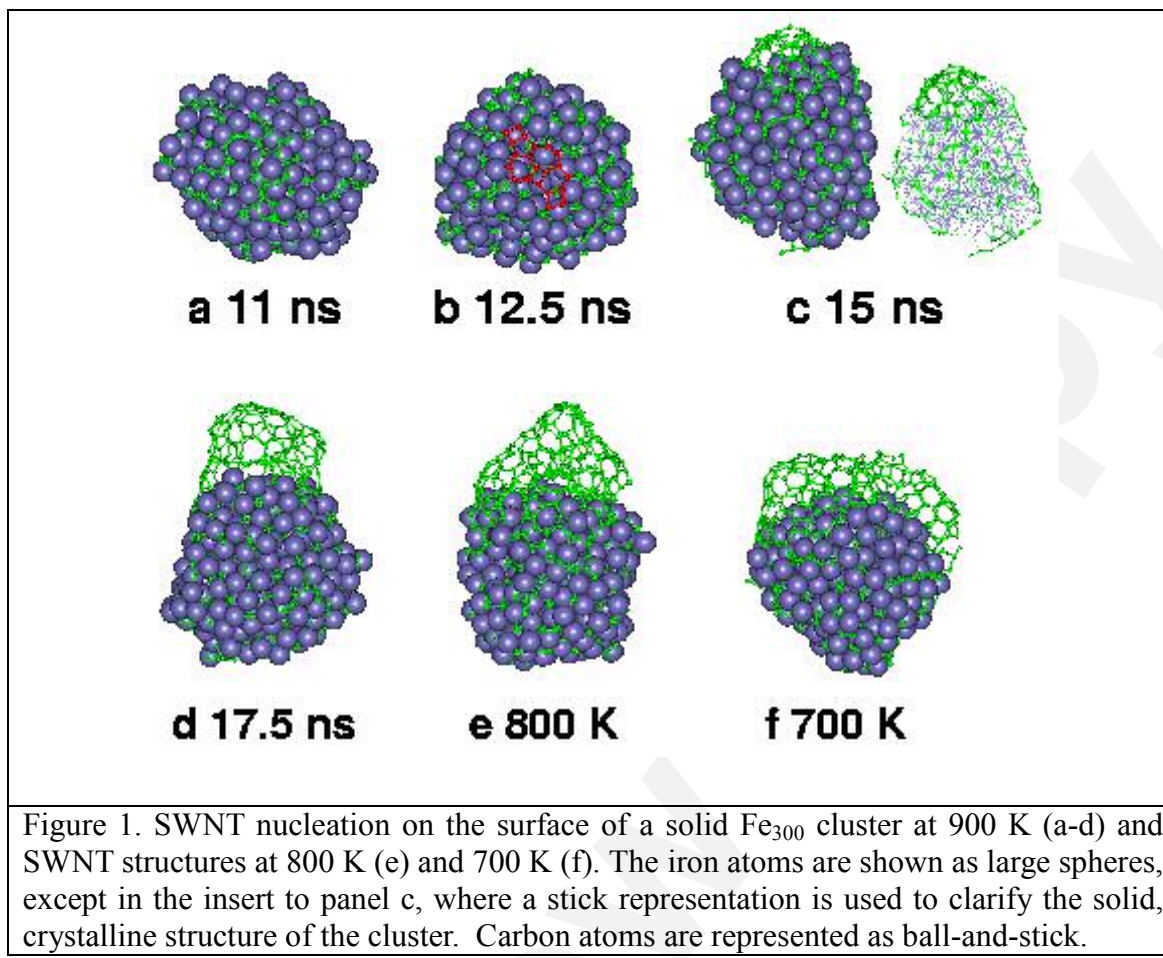

Figure 1. SWNT nucleation on the surface of a solid $Fe_{300}$ cluster at 900 K (a-d) and SWNT structures at 800 K (e) and 700 K (f). The iron atoms are shown as large spheres, except in the insert to panel c, where a stick representation is used to clarify the solid, crystalline structure of the cluster. Carbon atoms are represented as ball-and-stick.

Snap shots from a typical trajectory (using the first set of conditions) at 900 K is shown in Fig. 1a-d. This temperature is lower than the FeC cluster melting points, which are 1100 K, 950 K and 1000 K for $Fe_{300}$, $Fe_{300}C_{30}$ and $Fe_{300}C_{60}$, respectively. Many aspects of SWNT nucleation from the solid clusters, as exemplified in the figure, are similar to nucleation from liquid particles. For example, the C atoms on the cluster surface form carbon strings (Fig. 1a), which nucleate the formation of a graphitic island (Fig. 1b). The graphitic island grows in size and, at this temperature, the kinetic energy is sufficient to overcome the attraction between the island and the cluster, and the island lifts off the cluster to form the graphitic cap (Fig. 1c). This cap grows in diameter and length (Fig. 1d) to form the SWNT. Although the surface atoms of the cluster are not in their lattice positions (due to surface diffusion discussed below), the stick representation of the cluster atoms used in Fig. 1c clearly shows the crystalline structure of the solid FeC cluster at this temperature. SWNTs are also nucleated at even lower temperatures, as shown in Figs 1e and f for 800 and 700 K, respectively. [14]

The major difference between SWNT growth from solid and liquid particles is the diffusion path of the C atoms to the end of the growing SWNT. To illustrate this, the distance of each C atom from the cluster center of mass was monitored from the time that it was deposited on the cluster surface until it was incorporated in the SWNT wall, and the minimum distance was recorded. The insert to Fig. 2 shows a typical change in the distance of a C atom from the cluster center of mass after deposition on the cluster surface at 900 and 1200 K. It is clear that there is greater penetration of the C atom into the liquid cluster (1200 K) than into the solid cluster (900 K). The histogram in Fig. 2 is obtained from monitoring 157 C atoms and shows the number of C atoms as a function of their minimum distance from the cluster center of mass. For the liquid cluster (1200 K), many C atoms diffuse to the center of the cluster before incorporating into the SWNT structure, whereas there is far less penetration into the solid cluster (900 K) where the C atoms diffuse on (or near) the surface before being incorporated into the SWNT.

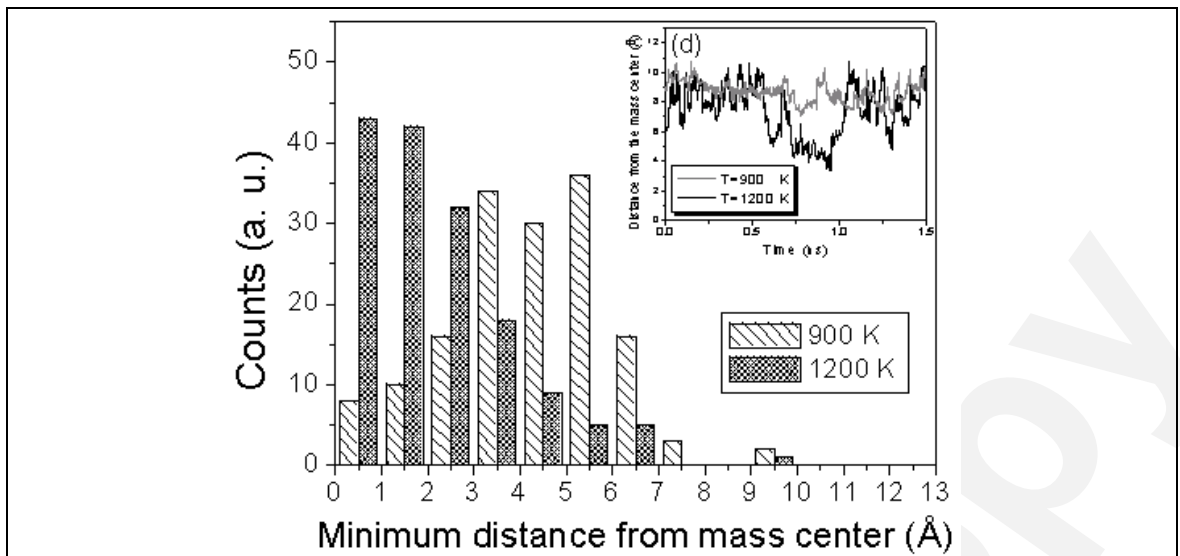

**Figure 2.** Histogram of the minimum distance of C atoms from the cluster center of mass before incorporation into the growing SWNT. The insert shows typical time dependence of the distance of the carbon atoms from the center of mass.

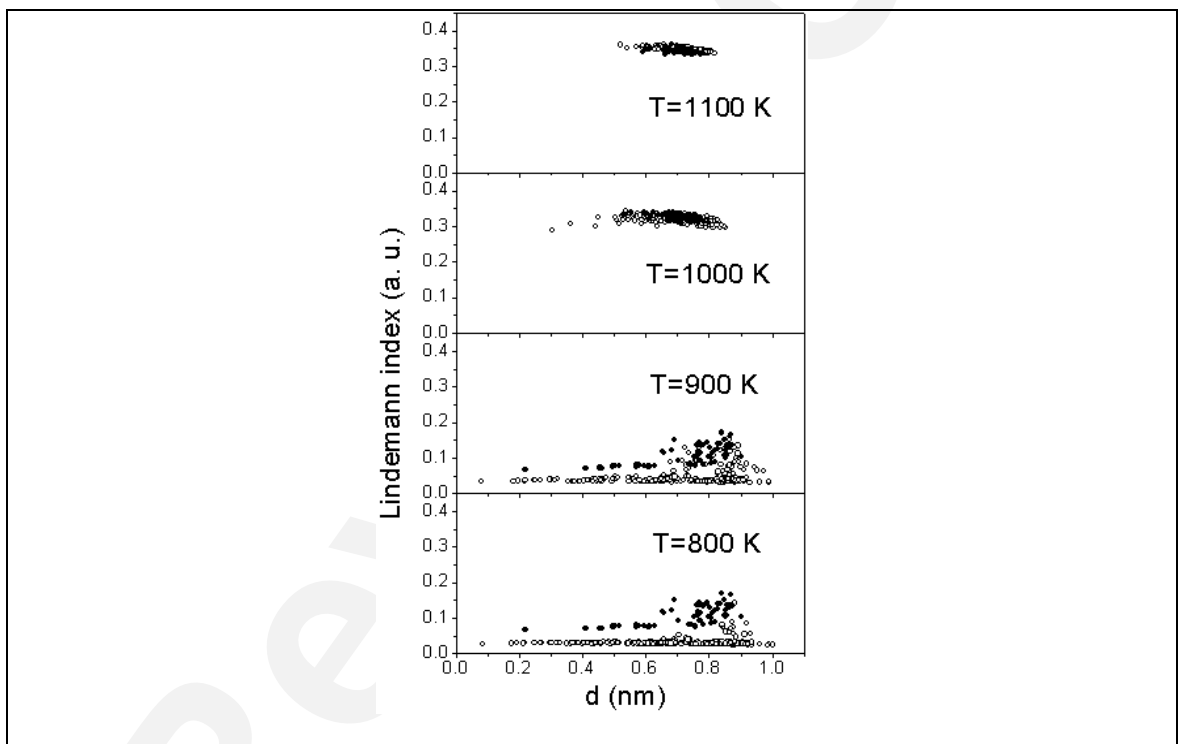

**Figure 3.** Lindemann indexes of the Fe (open circles) and carbon (solid circles) atoms in the $Fe_{300}C_{60}$ cluster as a function of their average distance from the cluster center of mass, d.

As noted with reference to Fig. 1a-d, the surface atoms in these 'solid' particles do not remain in their lattice positions during SWNT growth. This is observed in experiments as a continual change in shape of the catalyst particle during SWNT growth. [2, 15] Fig. 3 shows the Lindemann index for each atom in a $Fe_{300}C_{60}$ cluster as a function of its average distance from the cluster center of mass at 1100, 1000, 900 and 800 K. It is evident that carbon and iron atoms in the center of the FeC cluster remain in their lattice positions at 800 and 900 K, whereas the surface atoms are mobile. Thus, even at temperatures below the melting point the cluster surface has liquid-like features and C atoms can diffuse to the end of the growing SWNT. At 1000 and 1100 K the FeC cluster is liquid, and diffusion of atoms in to and out of the center of the FeC cluster is rapid.

In summary, MD simulations reveal that many aspects of SWNT nucleation and growth from solid and liquid metal particles are similar. In both cases graphitic islands lift off the cluster surface to form caps that grow into SWNTs. However, in contrast to liquid particles, where C atoms primarily diffuse into the bulk of the cluster before adding to the growing SWNT, incorporation of C into SWNTs on solid particles occurs predominantly via surface diffusion.

**Acknowledgement**

We are grateful to the Swedish Foundation for Strategic Research (CARAMEL consortium) and the Swedish Research Council for financial support, and for time allocated on the Swedish National Supercomputing facilities.

[14] Defects in the simulated SWNT are probably due to (a) the longer experimental times allow for thermal annealing of the SWNT, (b) the slower C precipitation in experiments leads to fewer defects at the SWNT end, and / or (c) catalytic healing of defects is not correctly described in the PES.

[15] Ichihashi T, Fujita J, Ishida M, Ochiai Y. In situ Observation of Carbon-Nanopillar Tubulization Caused by Liquidlike Iron Particles. Phys Rev Lett 2004;92:215702-1--215702-1.

8